\documentclass[11pt,twocolumn]{IEEEtran}
\usepackage{amsmath}
\usepackage{amssymb}
\usepackage{mathrsfs}
\usepackage{cite}
\usepackage{epsfig}
\usepackage{epsf}
\usepackage{theorem}
\usepackage{graphics}
\newenvironment{Proof}{\textit{Proof:}}{$\blacksquare$}

\newtheorem{thm}{Theorem}

\newtheorem{cor}{Corollary}

\begin{document}

\title{Limits on the Robustness of MIMO Joint Source-Channel Codes}
\author{
Mahmoud Taherzadeh and H. Vincent Poor
\\
\small Department of Electrical Engineering,
Princeton University,\\ Engineering Quadrangle, Olden Street
Princeton, NJ 08544 \\\small
e-mail: \{mtaherza, poor\}@princeton.edu
\thanks{This research was supported in part by the Air Force Office
of Scientific Research under Grant FA9550-08-1-0480.}
}
\maketitle \thispagestyle{empty}

\begin{abstract}
In this paper, the theoretical limits on the robustness of MIMO
joint source channel codes is investigated. The case in
which a single joint source channel code is used for the entire
range of SNRs and for all levels of required fidelity is considered. Limits on the
asymptotic performance of such a system are characterized in terms
of upper bounds on the diversity-fidelity tradeoff, which can be
viewed as an analog version of the diversity-multiplexing tradeoff.
In particular, it is shown that there is a considerable gap between
the diversity-fidelity tradeoff of robust joint source-channel codes
and the optimum tradeoff (without the constraint of robustness).
\end{abstract}

\section{Introduction}

Many applications call for the transmission of analog sources over
wireless channels. Results of research during the past decade have
shown that using multiple-antenna  systems can substantially improve
the rate and the reliability of communications in wireless fading
environments. Most research on multiple-antenna systems has focused
on the transmission of digital data over multiple-input
multiple-output (MIMO) channels, and the study of analog source
transmission over such channels is still in its early stages. In \cite{GunduzErkip},
\cite{CaireNarayanan} and \cite{BNCaire2006} some digital and hybrid digital-analog
techniques are examined for joint source-channel coding over MIMO
channels, and some bounds on the asymptotic exponents of the average
distortion are presented. In \cite{TaherzadehAsilomar08}, the
asymptotic exponents of the probability of having a large distortion
is studied. This measure, which is called the
\textit{diversity-fidelity tradeoff}, can be seen as an analog
version of the well-known diversity-multiplexing tradeoff which has
proven to be very useful in evaluating various digital space-time
coding schemes. In \cite{TaherzadehAsilomar08}, also some
semi-robust joint-source channel codes were proposed which can use
the same joint source-channel mapping for different ranges of SNR
and different ranges of desired resolution. However it was observed that
there is a gap between the optimum diversity-fidelity tradeoff and
the performance of those semi-robust codes. In this paper, we
investigate bounds on the robustness of MIMO joint
source-channel codes.

\section{System Model}

We consider a communication system in which an analog source of
Gaussian independent samples with variance $\sigma_{s}^{2}$ is to be
transmitted over an $(N_{t},N_{r})$ block fading MIMO channel where
$N_{t}$ and $N_{r}$ are the number of transmit and receive antennas
respectively. Each sequence of $m$ samples of the source,
represented by a vector $\mathbf{x}_{s}$, is transmitted over $n$
channel uses. We assume a quasi-static fading channel in which the
channel matrix $\mathbf{H}$ is fixed during these $n$ channel uses
and changes independently for the next $n$ channel uses. We call the
ratio $\eta=\frac{n}{m}$ the expansion/contraction factor of the
system. In a general setting, the communication strategy consists of
source/channel coding and source/channel decoding. As a result of
source channel coding, $\mathbf{x}_{s}$ is mapped into an
$N_{t}\times n$ space-time matrix $\mathbf{X}$ which in turn is
received at the receiver side as an $N_{r}\times n$ matrix
$\mathbf{Y}$ given by

$$\mathbf{Y}=\mathbf{HX}+ \sqrt{\frac{N_{t}}{\mathrm{SNR}}} \mathbf{W}$$
in which $\mathrm{SNR}$ is the average signal to noise ratio at each
receive antenna, and $\mathbf{W}$ is the normalized additive noise
matrix at the receiver whose entries are taken to be
$\mathcal{CN}(0,1)$ (the real variance of the noise is $\sigma^{2}=
\frac{N_{t}}{\mathrm{SNR}} $). At the receiver side, the
source/channel decoder yields an estimate of $\mathbf{x}_{s}$ from
$\mathbf{Y}$ as $\widehat{\mathbf{x}_{s}}$. For a specific channel
realization $\mathbf{H}$, the distortion measure is

\begin{equation}
D(H)=\mbox{E}_{\mathbf{x}_{s}}\{\|\mathbf{x}_{s}-\widehat{\mathbf{x}_{s}}\|^{2}\vert \mathbf{H}\}.
\end{equation}

For any specific strategy, we define the $f-$fidelity event as
$\mathcal{A}(f)=\{\mathbf{H}: D(\mathbf{H})>\mathrm{SNR}^{-f}\}$ and
we call $f$ the fidelity exponent. For specific values of $\eta$,
$N_{t}$ and $N_{r}$, we define

\begin{equation}
d(f)=\lim_{\mathrm{SNR}\rightarrow{\infty}}-\frac{\log \Pr \{
\mathcal{A}(f)\}}{\log\mathrm{SNR}}.
\end{equation}

We call $d(f)$ the diversity, and denote its maximum (over all possible source-channel coding schemes) as $d^{\ast}(f)$.

In \cite{TaherzadehAsilomar08}, it is shown that the optimal diversity (if we can use different source-channel codes for different SNR values and different fidelity exponents) can be characterized as
\begin{equation}
d^{\ast}(f)= \left( N_{t}-\frac{f}{2\eta} \right) \left(
N_{r}-\frac{f}{2\eta} \right)
\end{equation}
for integer values of $\frac{f}{2\eta} $.

\section{Bounds on the diversity-fidelity tradeoff of a single MIMO source-channel map}

In this paper we investigate upper bounds on the diversity-fidelity
tradeoff, when the joint source-channel code is fixed. In the
general case, this joint source-channel code is a mapping from the set
of all $m$-tuples of source samples to $\mathcal{F}$, the set of
transmitted vectors (or the modulation set), which is a subset of
$\mathbb{R}^{2nN_{t}}$ (or indeed $\mathbb{C}^{nN_{t}}$). We assume
that $N_{r}\geq N_{t} $. Also, we focus on the case in which the
source is uniformly distributed on $[0,1] $, which has variance
$\frac{1}{12}$.

To obtain bounds on the diversity-fidelity tradeoff of a single MIMO
source-channel map, we use the concept of \emph{box-counting
dimension} \cite{EdgarFractal}. If we partition the space into a
grid of cubic boxes of size $\sigma$, and consider $ N_{\sigma}$ as
the number of boxes that intersect the set $\mathcal{F}$, the
box-counting dimension of $\mathcal{F}$ is defined as

\begin{equation}{\rm Dim}(\mathcal{F}) \triangleq \lim_{\sigma \rightarrow 0} \frac{\log N_{\sigma}}{\log\frac{1}{\sigma}}.\end{equation}

We modify this definition and define the \emph{$c$-effective box-counting dimension} (for $0 < c \leq 1 $) of a modulation set as
\begin{equation}{\rm Dim}_{c}(\mathcal{F}) \triangleq \lim_{\sigma \rightarrow 0} \frac{\log N^{\prime}_{c, \sigma}}{\log\frac{1}{\sigma}}
\end{equation}
where $N^{\prime}_{c, \sigma}$ is the minimum number of those boxes whose total probability of containing the modulated signal is at least $c$.

\begin{thm}
Consider a space-time joint source-channel coding with modulation
set $\mathcal{F}$ (mapping $m$-dimensional source vectors to
$2nN_t$-dimensional transmitted vectors). If for every $c>0$, the
$c$-effective box-counting dimension of $\mathcal{F}$ is at least $2
n \beta$ and at most $2n \beta^{\prime}$, then for any $0\leq f \leq
2\eta \beta^{\prime}$, we have
\begin{equation}
d(f)\leq \left( N_{r}-\beta+1  \right) \left(N_{t}-\beta+1
\right)\left(1-\frac{f}{2 \eta \beta^{\prime}} \right).
\end{equation}
\end{thm}

\begin{Proof}
For any positive numbers $0< c_{1}<c_{2}<1$, if ${\rm
Dim}_{c_1}(\mathcal{F})=\beta_1$ and ${\rm
Dim}_{c_2}(\mathcal{F})=\beta_2$, then $2n \beta^{\prime} \geq
\beta_1 \geq \beta_2 \geq 2n \beta$, and for any $\sigma$ and for
any of the boxes corresponding to $c_1$, the probability of
containing the modulated signal is at least in the order of
$\sigma^{\beta_2} $ and their number is of the order of $
\sigma^{-\beta_1}$. Based on the monotonicity of ${\rm
Dim}_{c}(\mathcal{F})$, we can find $c_1$ and $ c_2$ such that
$\beta_1 $ and $ \beta_2$ are arbitrarily close to each other.

Now we look at the received modulation set $\mathbf{H}\mathcal{F}$.
We denote the nonzero eigenvalues of
$\mathbf{H}\mathbf{H}^{\mathtt{H}}$ by $0<\lambda_{1} \leq ... \leq
\lambda_{N_t}$, and consider $\alpha_{i}= \frac{\log
\lambda_{i}}{\log \sigma }$. If $\alpha_{i}\geq 1$ (for $1 \leq i
\leq N_t - \beta +1$), then $\mathbf{H}\mathcal{F}$ (and all the
boxes corresponding to $c_1$) will be inside a $2nN_{t}
$-dimensional orthotope whose volume is less than
$\sigma^{2n(N_{t}+1-\beta)}$. In this case, because the order of the
number of the boxes corresponding to $c_1$ is greater than $\sigma^{
-2n\beta}$, the majority of them (with their portion approaching to
1) become adjacent to other boxes (with a distance less than
$\sigma$). Also, because of the isotropy of the channel distribution
and the eigenvectors of $ \mathbf{H}$, with probability
approaching to 1, this also includes boxes containing the mapping of
distant sub-segments of the source. Therefore, in this case, the
distortion becomes lower bounded by a positive number, and hence the
fidelity exponent will be $f=0$.

Thus, to bound $d(0)$ we need only to bound $\Pr \left\{
\alpha_{i}\geq 1 \vert 1 \leq i \leq N_t - \beta +1 \right\}$.
Similarly to \cite{ZT2003}, we can bound it as \footnote{In this
paper, we use $a\doteq b$ to denote that $a$ and $b$ are
asymptotically equivalent.}
$$\Pr \left\{\alpha_{i}\geq 1 \vert 1 \leq i \leq N_t - \beta +1 \right\}  \doteq $$ 
$${\rm SNR}^{-(N_{t}-\beta+1)(N_{r}- \beta+1)}$$
\begin{equation}
\Rightarrow d(0) \leq  (N_{t}-\beta+1)(N_{r}-\beta+1).
\end{equation}

For $f>0$, we use a similar approach, by considering the effect of
the channel on the boxes of size $\sigma$ inside larger boxes of
size $\sigma^{\frac{f}{2\beta^{\prime} \eta}} $ (containing at
least approximately $\sigma^{-\frac{f}{2}} $ smaller boxes). Consider
$\alpha^{\prime}= \frac{\log \lambda_{i}}{\log \sigma^{\left(1-
\frac{f}{2 \beta^{\prime} \eta}\right) }} $. Now if
$\alpha^{\prime}_{i}\geq 1$ (for $1 \leq i \leq N_t - \beta +1$),
similarly to the case of $ f=0$, we can show that the distortion
will be at least on the order of $\sigma^{2f}$ (or ${\rm
SNR}^{-f}$). Therefore, we have
\begin{equation}
d(f)\leq \left( N_{r}-\beta+1  \right) \left(N_{t}-\beta+1
\right)\left(1-\frac{f}{2 \eta \beta^{\prime}} \right).
\end{equation}
\end{Proof}

\begin{thm}
Consider a space-time joint source-channel coding with modulation
set $\mathcal{F}$ (mapping $m$-dimensional source vectors to
$2nN_t$-dimensional transmitted vectors). If for some $c>0$ the
$c$-effective box-counting dimension of $\mathcal{F}$ is at most $2n
\beta$, then the coding scheme cannot achieve a fidelity exponent
larger than $2 \eta \beta$.
\end{thm}

\begin{Proof}
If we divide the signal space into boxes of size $\sigma$, the
number of boxes corresponding to $c$ is bounded by the order of
$\sigma^{-2 n \beta}$. If we divide the subset of source vectors
(that are mapped into these boxes) to $ \sigma^{2n(-\beta-
\varepsilon)}$ sub-cubes whose size is on the order of
$\sigma^{2\eta (\beta+\varepsilon)} $, a large portion of them will
be adjacent to each other (with a distance less than $\sigma$),
hence $\Pr \left\{D > {\rm SNR}^{-2\eta (\beta+\varepsilon)}
\right\} $ can be lower bounded by a positive number. This argument
is valid for any small $\varepsilon$. Thus, the fidelity exponent
cannot be larger than $2 \eta \beta$.
\end{Proof}

\begin{cor}
No single joint source-channel mapping can achieve any point on the optimum diversity-fidelity curve, other than the two extreme points, $d=0$ or $f=0$.
\end{cor}

Theorems 1 and 2 show that the effective dimensionality of the
analog modulation set is a key factor in determining its asymptotic
performance. While low-dimensional mappings are incapable of
achieving a high fidelity exponent, high-dimensional mappings
cannot achieve a high diversity order. This is totally different
from the case of digital space-time coding, in which many full-rank
lattice codes can be used to construct diversity-multiplexing-tradeoff-achieving space-time
codes (assuming that maximum-likelihood decoding is performed at the
receiver).

\begin{figure}
  \centering
  \includegraphics[scale=.5,clip]{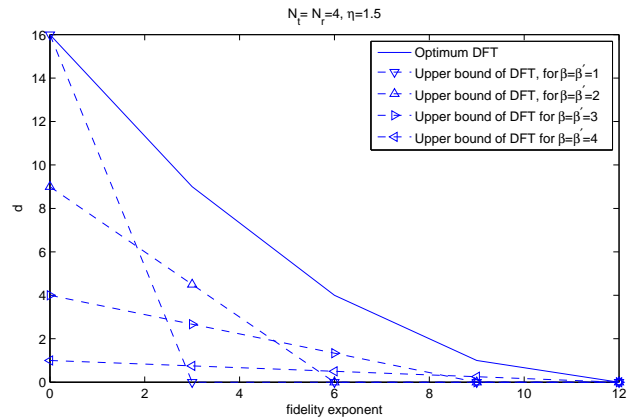}

  \caption{Bounds on the diversity-fidelity tradeoff of robust codes (depending on their effective dimensions) for 4 transmit and 4 receive antennas and bandwidth expansion $ \frac{3}{2}$.}
  \label{}
\end{figure}


\section{conclusions}
In this paper, we have introduced an upper bound on the
diversity-fidelity tradeoff of single-mapping MIMO source-channel
codes. This result shows that, unlike the case of a
single-input/single-output (SISO) channel (in which we can achieve
the optimum signal-to-distortion-ratio (SDR) scaling by using a single mapping
\cite{ISIT-analog}), in the MIMO case there is a considerable gap
between the asymptotic performance of a single robust mapping and
the optimum tradeoff.

\bibliography{thesis}

\newcommand{\noopsort}[1]{} \newcommand{\printfirst}[2]{#1}
  \newcommand{\singleletter}[1]{#1} \newcommand{\switchargs}[2]{#2#1}
\begin{thebibliography}{1}

\bibitem{GunduzErkip}
D.~G\"{u}nd\"{u}z and E.~Erkip, ``Distortion exponents of {MIMO} fading
  channels,'' in {\em Proceeding of the IEEE Information Theory Workshop},
  (Punta del Este, Uruguay), pp.~694 -- 698, July 2006.

\bibitem{CaireNarayanan}
G.~Caire and K.~Narayanan, ``On the distortion {SNR} exponent of hybrid digital
  – analog space – time coding,'' {\em IEEE Trans. Inf. Theory}, vol.~53,
  pp.~2867 -- 2878, August 2007.

\bibitem{BNCaire2006}
K.~Bhattad, K.~Narayanan, and G.~Caire, ``On the distortion exponent of some
  layered transmission schemes,'' in {\em Proceedings of the Asilomar
  Conference on Signals, Systems and Computers}, (Pacific Grove, CA), Nov.
  2006.

\bibitem{TaherzadehAsilomar08}
M.~Taherzadeh and A.~K. Khandani, ``Diversity-fidelity tradeoff in transmission
  of analog sources over {MIMO} fading channels,'' in {\em Proceedings of the
  Asilomar Conference on Signals, Systems and Computers}, (Pacific Grove, CA),
  Oct. 2008.

\bibitem{EdgarFractal}
G.~A. Edgar, {\em Measure, Topology and Fractal Geometry}, ch.~6.
\newblock Berlin: Springer-Verlag, 1990.

\bibitem{ZT2003}
L.~Zheng and D.~Tse, ``Diversity and multiplexing: {A} fundamental tradeoff in
  multiple-antenna channels,'' {\em IEEE Trans. Info. Theory}, pp.~1073--1096,
  May 2003.

\bibitem{ISIT-analog}
M.~Taherzadeh and A.~K. Khandani, ``Robust joint source-channel coding for
  delay-limited applications,'' in {\em Proceedings of the 2007 IEEE
  International Symposium on Information Theory}, (Nice, France), June 2007.

\end{thebibliography}
\bibliographystyle{ieeetr}

\end{document}